# Precipitation of corrosion products in macroscopic voids at the steel-concrete interface – observations, mechanisms and research needs

Shishir Mundra[1,a], Emanuele Rossi[1], Luka Malenica[1], Mohit Pundir[1], Ueli M. Angst[1]

[1]Institute for Building Materials, ETH Zürich, Laura-Hezner-Weg 7, 8093, Zürich, Switzerland

Corresponding author: [a] smundra@ethz.ch

## Abstract

Macroscopic voids at the steel-concrete interface and their degree of saturation with an aqueous electrolyte are known to play an important role in the corrosion of steel in reinforced concrete. Irrespective of the exposure conditions and testing parameters, corrosion products have been reported to consistently precipitate in a unique pattern within these macroscopic voids, preferentially along the void walls and growing inward. The underlying mechanisms governing corrosion product precipitation in macroscopic voids and their effects on long-term durability remain unclear. Through in-situ X-ray computed tomography observations, thermodynamic and kinetic considerations, and numerical modelling of water transport within macroscopic voids, here, we provide plausible hypotheses of the processes responsible for the precipitation of corrosion products along the walls of the voids. Understanding the mechanisms of corrosion product precipitation can offer insights into the development of stresses in and around the macroscopic interfacial void and the durability of reinforced concrete structures. This contribution also discusses opportunities for different avenues for research to elucidate several multiscale processes that influence the durability of reinforced concrete.

## 1. Introduction

The corrosion of steel in concrete represents a very complex case of metallic corrosion in a reactive porous media, primarily because the steel-concrete interface (SCI) is heterogeneous both in microstructure as well as in chemistry spatially and temporally; and can be considerably distinctive in different structures [1–4]. The inhomogeneities at the SCI and the complexity of the microscopic physical-chemical processes occurring in corroding reinforced concrete structures can have multi-scale effects on the durability of concrete structures [5].

In this context, macroscopic or coarse voids at the SCI and their degree of saturation with aqueous electrolyte, were found to be important characteristics that could influence both corrosion initiation and propagation [6]. Macroscopic voids in concrete, defined as voids with diameters ranging from tens of microns to a few mm, are inevitably found at the SCI as they form when air bubbles become entrapped and entrained at the SCI during the compaction of fresh concrete. The moisture content within these voids (completely air-filled, partially filled with solution, or completely saturated) may vary over time as a function of the surrounding environmental conditions. Depending on the moisture and oxygen conditions of interfacial voids, several mechanisms pertaining to the initiation of corrosion w.r.t. the location of the interfacial void have been hypothesised and discussed by RILEM TC-262 SCI [6]. For a deeper understanding of the influence of moisture in macroscopic interfacial voids on corrosion initiation, we recommend referring to the critical review by RILEM TC-262 SCI [6].

While the focus in the literature has predominantly been on corrosion initiation, there has been comparatively less emphasis on understanding the propagation phase and precipitation of corrosion products in the concrete matrix. In this context, it particularly noteworthy that corrosion products generally precipitate in a specific pattern along the walls of macroscopic voids at the SCI, as will be presented in detail in Section 2. Though such an observation may have several implications on transport behaviour and on our understanding of the processes occurring between the initiation of corrosion and the end of serviceability of the structure, we currently lack an explanation as to why and how corrosion products precipitate in this particular fashion and its consequences on durability. In this manuscript, we aim to provide a conceptual explanation for this behaviour based on recent in-situ experimental observations, theoretical thermodynamic and kinetic considerations and numerical modelling of water transport within these macroscopic interfacial voids. Additionally, understanding the underlying mechanism governing the precipitation of corrosion products within interfacial defects, such as macroscopic voids, may help us decipher implications on how fracture of the surrounding concrete matrix occurs.

## 2. Experimental observation of precipitation of corrosion products in voids

## 2.1. *Observations in the literature*

Despite inconsistent experimental results [7–18], it is common to witness the occurrence of corrosion initiation either within macroscopic voids at the SCI or in their vicinity, with corrosion products precipitating within the void and spreading in the surrounding areas [19]. These experimental findings are often derived from two dimensional cross sections of samples studied at a particular time using scanning electron microscopy (SEM) (as illustrated in Figure 1). The destructive nature of sampling of the SCI for SEM imaging may induce artefacts such as cracks, debonding etc. and lead to misinterpretation of the findings [19]. Nevertheless, irrespective of exposure conditions and other testing parameters, the precipitation of corrosion products within the interfacial macroscopic void was in many studies reproducibly found to occur in a peculiar and consistent pattern (Figure 1). In particular, the precipitation of corrosion products occurs along the walls of interfacial voids in a layered fashion and with each layer being of a different composition (as visible from backscattered SEM images in Figure 1 and marked with yellow arrows). From the two-dimensional images in Figure 1, the corroding zone is often (partially) encompassed by the macroscopic void, and the precipitation of corrosion products is more or less concurrent along the walls of the void, both close to and away from the corroding zone. For example, the backscaterred SEM micrograph shown in Figure 1d represents a corroding section of a reinforced concrete specimen, where distinct layers of different corrosion products (different textures and grey-scale intensities) can be seen all along the walls of a macroscopic interfacial void roughly two weeks after corrosion initiation.

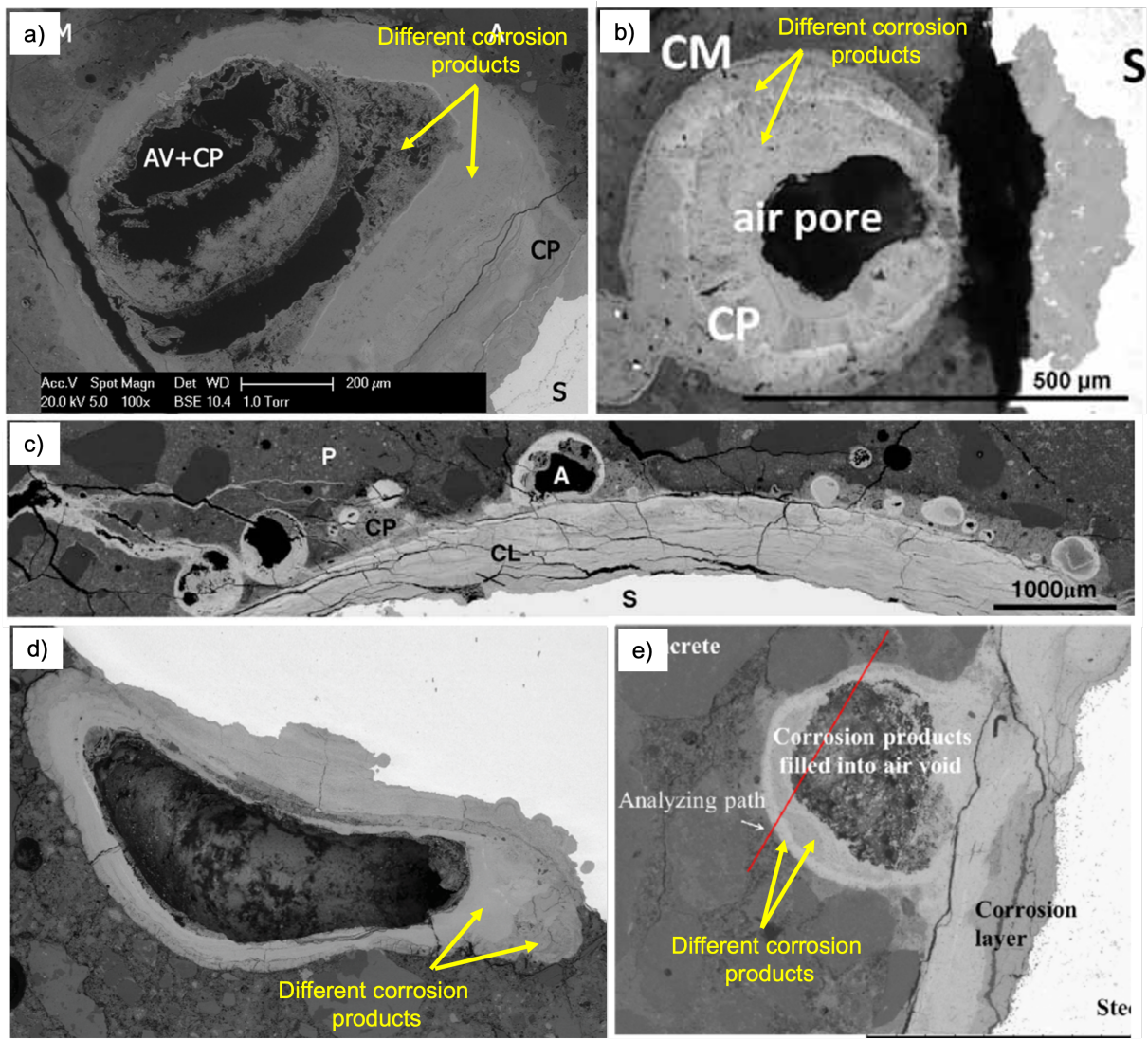

Figure 1: Scanning electron micrographs (adapted) from the literature (reprinted from [7,8,20–22] with permission from Elsevier and John Wiley & Sons) of macroscopic voids at the SCI and near the corroding zone, with layers of different corrosion products precipitated along the surface of the void (marked with yellow arrows). Irrespective of the exposure conditions and other testing parameters, the majority of micrographs indicate a consistent pattern of precipitation along the surface of macroscopic interfacial voids while the inner region of the void remains empty.

The pitfalls of relying on destructive tests [19,23] to study the precipitation of corrosion products have partially been overcome through the use of advanced non-destructive imaging techniques such as X-ray computed tomography (XCT), allowing in-situ imaging of multi-scale processes over time [24]. Additionally, the use of neutron radiography/tomography (either independently or in conjunction with XCT) has allowed to investigate the influence of the moisture content locally within macroscopic interfacial voids on corrosion initiation [20,24]. XCT is now widely used in studies investigating localised corrosion, crack growth propagation, precipitation of corrosion products etc. [24–27]. The main limitations of XCT include the size of the sample, and the spatial and temporal resolution that can be achieved [28].

While ex-situ SEM (also in combination with other characterising techniques such as Raman spectroscopy and/or EDX – as used in [29,30]) indicates the presence of different corrosion products along the walls of the void, we currently lack insights into how these corrosion products precipitate over time. Several authors have indicated, through XCT, a similar precipitation pattern (like those shown in Figure 1) within macroscopic interfacial voids. Bernachy-Barbe et al. [31] pointed out in their XCT study, corrosion products first precipitate on the walls of macroscopic interfacial voids and their growth over time occurs inwards towards the centre of the void, rather than outwards into the concrete matrix (Figure 2a). However, the similarities in the linear attenuation coefficients of different corrosion products does not allow us to distinguish their layered structure and differences in composition. Several other studies have also reported similar precipitation behaviour either under naturally corroding conditions [8] or under artificially accelerated (galvanostatically induced) corrosion [24–27]. It is to be noted that precipitation of corrosion products in macroscopic interfacial voids may not always exhibit the pattern highlighted in Figure 1, as shown in the neutron tomograph obtained by Robuschi et al. [24] (Figure 2b). However, the focus of this manuscript is primarily on the most commonly observed behaviour of precipitation (Figure 1) in macroscopic interfacial voids (like those observed by Bernachy-Barbe et al. [31] in Figure 2a).

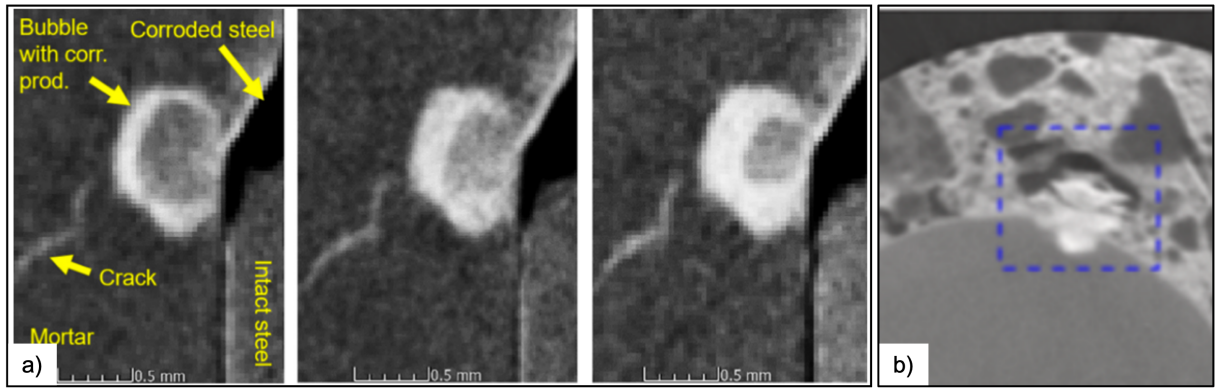

Figure 2: (a) X-ray and (b) neutron computed tomographs from the literature showing the precipitation of corrosion products within macroscopic voids at the SCI. Continual in-situ XCT observations (Figure 2a) by Bernachy-Barbe et al. [31] (reprinted with permission from Elsevier) on cracked specimens, suggest that corrosion products first precipitate on the walls of the interfacial void and with time, grow inwards. In some cases (like in Figure 2b), Robuschi et al. [24] observed corrosion products to not form along the walls of the void, instead precipitate within the corroding zone and its vicinity (indicated by the dashed blue square).

## 2.2. *X-ray CT measurements on corroding reinforced mortars (this study)*

In this study, we make use of XCT[1] to understand and monitor the precipitation of corrosion products at the SCI and the moisture state within macroscopic interfacial voids, over time, in a reinforced mortar specimen exposed to natural corroding conditions (Figure 3a-k). Post localised corrosion initiation within the macroscopic interfacial void (with diameter > 1 mm) or in its vicinity, the corroding zone (Figure 3c, d) reaches a depth of 400 μm in the time between 4 and 8 weeks of exposure to wet and dry cycles. The precipitation of corrosion products is clearly visible within the localised corroding zone, as well as at the walls of the interfacial void in the form of a μm-thick layer (i.e., outer layer (CP)). We would like to draw the attention of the reader to the fact that the observations in our experimental study (Figure 3) do not allow us to elucidate on the conditions prevailing between the onset of active corrosion and the precipitation of the outer layer of corrosion products on the walls of interfacial voids (Figure 3d).

---

[1] XCT data on reinforced concrete specimens, exposed to wet and dry cycles with a 3.5 wt.% NaCl solution, were acquired at different times at the ICON beamline in Paul Scherrer Institut (PSI, Villigen, Switzerland). Experimental details and methodology of XCT of reinforced concrete specimens can be found in the Supplementary Note 1.

Interestingly, during the drying phase of the wet and dry cycles (after 8 weeks of continuous exposure to wet and dry cycles), XCT acquisitions[2] revealed that this interfacial void remains filled with solution, while all the other macroscopic voids (Figure 3c) dispersed in the mortar matrix are filled with air. We infer the presence of solution in this interfacial void due to the significantly different (lower) X-ray linear attenuation coefficient of air in comparison to water (as highlighted in [32]). It is worthwhile to note that the solution visible inside the void (Figure 3c, d) does not necessarily resemble the moisture conditions at which corrosion initiated. This is due to the fact that the specimen was continually subjected to wet-dry cycles between XCT acquisitions, and the moisture conditions shown in Figure 3c, d may potentially be due to the wetting phase preceding the XCT acquisition. This topic will be treated in detail later.

Over time and subsequent wet-dry cycles, the precipitation of a second layer of corrosion products occurs, growing inwards from the walls of the void and thereby, reducing the volume of the void (Figure 3e-f). The same behaviour can be observed in a smaller void (diameter ~1 mm) along the SCI, at around 2 mm distance from the initial corroding zone (Figure 3f, section C-C'). Upon prolonged drying of the specimen in laboratory conditions (from Figure 3e-f to Figure 3g-h), the solution inside both interfacial voids dries out and gets replaced by air, and no further precipitation of corrosion products occurs. It is also interesting to notice that, after an initially aggressive growth of the corroding zone on the steel (Figure 3c-d), the corroding zone does not significantly increase in depth over time but spreads laterally, along the width and length of the steel reinforcement (Figure 3i, k). As a consequence, precipitation of corrosion products is not localised in the proximity of the initial corroding spot, but also spreads along the steel surface (Figure 3f, h, i, k), and also commonly reported in the literature (Figure 1) [7,8,20–22].

---

[2] XCT acquisitions were always conducted during the drying phase of the wet and dry cycles, as explained in Supplementary Note 1.

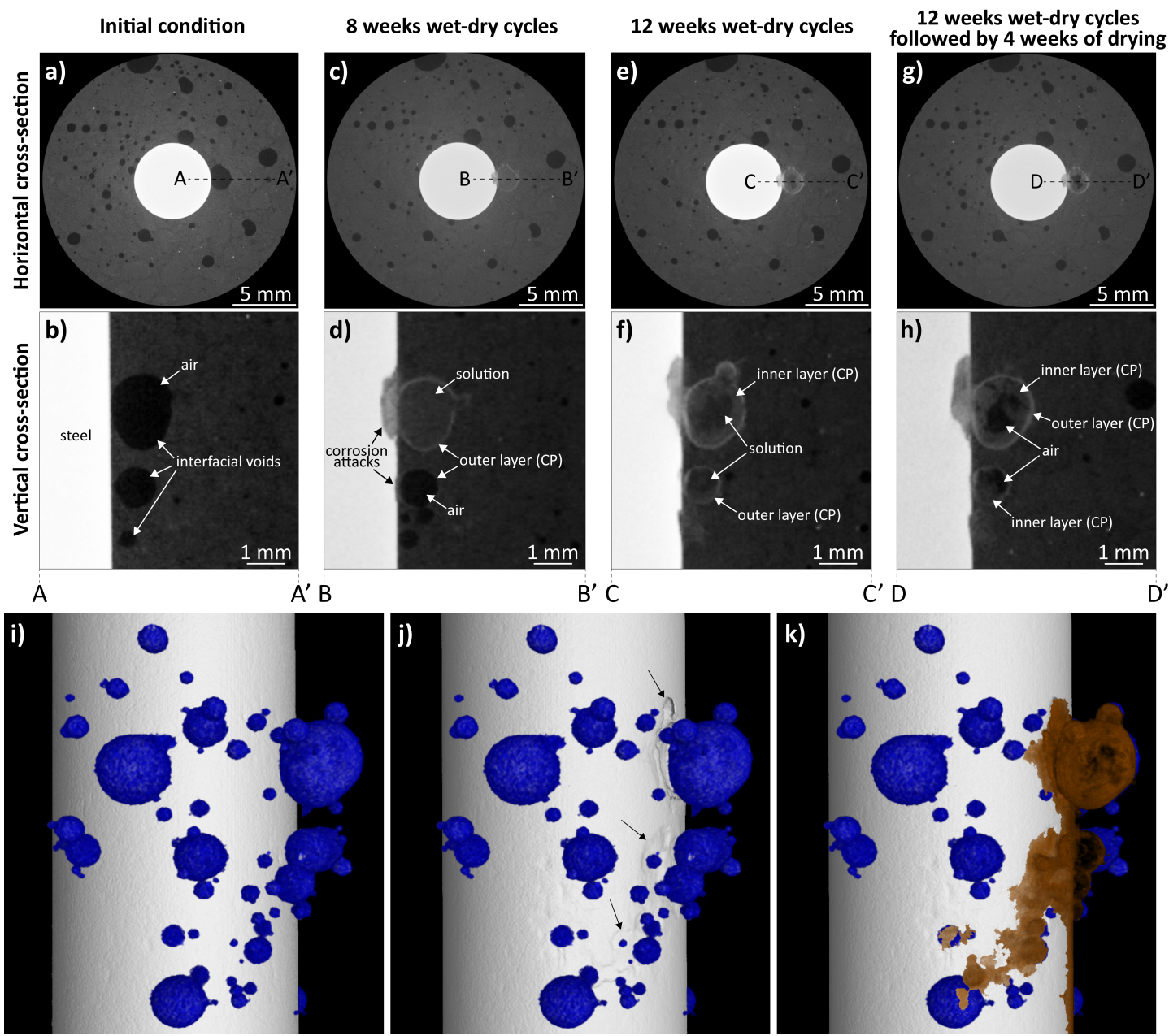

Figure 3: XCT 2D horizontal (a, c, e, g) and vertical (b, d, f, h) cross-sections obtained from a corroding reinforced mortar specimen at different times of exposure to wet and dry cycles. These in-situ XCT measurements (experimental details in Supplementary Note 1) clearly show the precipitation of corrosion products (CP) along the walls of the macroscopic void at the SCI and its inward growth over time. Figures i, j, and k show the XCT 3D renders of the corroding reinforced mortar sample in its initial condition (i), highlighting the interfacial void (in blue); and after undergoing 12 weeks of exposure to wet-dry cycles followed by 4 weeks of drying (j, k). While (j) indicates the spread of corrosion attacks along the steel surface (indicated by black arrows), (k) highlights the distribution of corrosion products (in brown)

within the void and along the steel surface. Air, solution and corrosion products within the interfacial voids are distinguished based on the grey scale values.

## *2.3. Gaps in understanding precipitation of corrosion products in macroscopic interfacial voids*

Based on ex-situ SEM and in-situ XCT experiments, there exists a broad consensus within the literature that precipitation of corrosion products occurs within the macroscopic voids at the SCI. Independent of the exposure conditions, both ex-situ and in-situ observations revealed a unique and reproducible pattern of precipitation of corrosion products. Corrosion products first precipitate along the walls of macroscopic interfacial voids, and then grow inwards as distinct layers and with different compositions over time. The results obtained from XCT measurements conducted in this study, in addition, provided insights into the moisture state of the macroscopic interfacial void. Once a layer of corrosion products forms along the walls of the void, the water within this void does not dry out as quickly as the other voids in the cementitious matrix during the drying phase of wet and dry cyclic exposure. Over subsequent exposure to wet and dry cycles, a second layer of corrosion products precipitates within the void, along the surface of the first layer of corrosion products. Only when the specimens are dried for a prolonged duration, the water within this void is replaced by air. Such observations, similarities and consistencies in multiple specimens exposed to a variety of environments make one wonder what processes govern the precipitation of different corrosion products along the walls of the macroscopic void at the SCI. Specifically, three major scientific questions arise from the observations shown above:

A) Why do corrosion products precipitate preferentially along the walls of macroscopic interfacial voids as a 'shell', and grow inwards in layers?
B) What factors lead to differences in the chemistry of corrosion products that have precipitated in different layers over time?
C) Why do macroscopic interfacial voids that exhibit this pattern of precipitation have a different moisture state when compared to other concrete voids (of similar geometry) without any corrosion products?

# 3. Mechanistic understanding of precipitation of corrosion products in macroscopic interfacial voids

In light of both ex-situ (Figure 1) and in-situ (Figure 2 and Figure 3) observations from this study and the literature, we take a holistic view (Figure 4) of the processes occurring from the time of sustained anodic dissolution of iron up to the precipitation of corrosion products along the periphery of a void. Here, we try to explain the multitude of processes that may be responsible for the observation of corrosion products precipitating at the walls of the void. Additionally, we also propose several research avenues that are essential to develop a thorough mechanistic know-how of the processes leading to the precipitation of corrosion products in macroscopic interfacial voids with a peculiar and consistent pattern (as evidenced in Figure 1, Figure 2, and Figure 3). It must be explicitly noted that the proceeding discussion pertains to cases where corrosion has occurred in the proximity of (or within) partially saturated macroscopic voids that may commonly be encountered at the SCI [32]; however, a similar hypothesis could most likely be applicable to fully saturated macroscopic voids at the SCI.

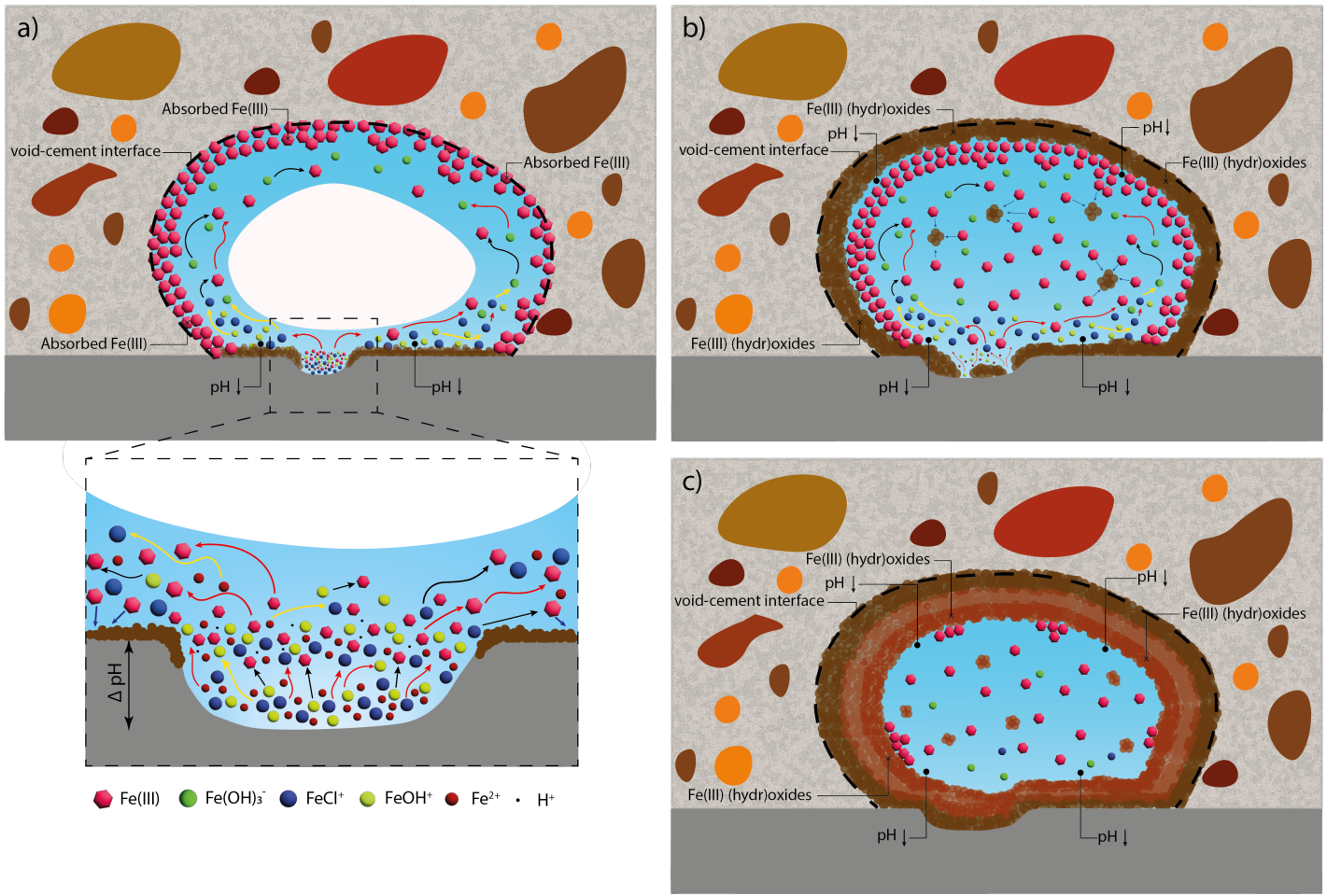

Figure 4: Schematic representation of the hypothesised mechanism of corrosion products precipitation occurring within partially saturated interfacial voids; a) diffusion of Fe ions (red arrows), oxidation (black arrows), hydrolysis (yellow arrows) from the active steel area, adsorption of Fe(III) at the walls of the void and precipitation of corrosion products (blue arrows); b) precipitation of the first layer of corrosion products at the walls of the void, reducing the desorption rate of solution from the void volume; c) inward growth of a second and subsequent layers of corrosion products, precipitated from the solution present inside the void.

*Processes leading up to the precipitation of the first layer of corrosion products*

Fe(II) and Fe(III) ions diffuse away from the mouth of the corroding zone into the electrolyte within the void. Given that sustained anodic dissolution would occur at a much lower local pH [33,34] than that of the concrete pore solution, it is reasonable to assume that the pH of the electrolyte within the void is heterogeneous. Specifically, the pH within the corroding zone is probably lower than the pH near or adjacent to its mouth and significantly lower than that encountered along the walls of the void (further away from the local corroding zone), where the pH is buffered (at ~ 12.5 to 13) by the cementitious phases. As Fe(II) ions diffuse out, the molar fraction of hydrolysed Fe(II) species (such as $Fe(OH)^+$, $Fe(OH)_2$ (aq.), and $Fe(OH)_3^-$) w.r.t. $Fe^{2+}$ increases as the pH of the electrolyte increases and consequentially, the rate at which they oxidise to Fe(III) also increases [35,36]. Therefore, the majority of the aqueous iron ions within the electrolyte present in the interfacial void would be in the +3-oxidation state but in different hydrolysed states – depending on the local pH of the electrolyte within the void. The solubility of Fe(III) is much lower than Fe(II) in the near-neutral pH range (pH: 6 – 9) and the precipitation of the Fe(III) (hydr)oxides would occur almost instantaneously, thereby reducing the local concentration of Fe(III) to the solubility limit of the precipitating Fe(III) (hydr)oxide [35]. Additionally, the precipitation of Fe(III) (hydr)oxides involves the generation of protons, and therefore, would reduce the local pH [37]. The amount of corrosion products precipitated in regions adjacent to the mouth of the corroding zone exceed that observed along the void-cement interface (Figure 1 and Figure 3d). This suggests that the chemistry of the electrolyte in regions adjacent to the mouth of the corroding zone could resemble those where the solubility of Fe(III) (hydr)oxides is approximately minimum, i.e. near-neutral pH conditions [35] (Figure 4a).

In regions where the pH > ~10 (i.e. further away from the mouth of the corroding zone), Fe(III) is more soluble than Fe(II) and is potentially more mobile [35]. Mancini et al. [38] showed that the ability of Fe(III) ions to sorb onto C-S-H or get incorporated into other cement hydrate phases at alkaline conditions [39] is significant. Therefore, upon diffusion of Fe(III) in the electrolyte (along the walls of the void, as shown in Figure 4a) within the interfacial void, Fe(III) ions preferentially sorb onto the void-cement interface. Depending on the local pH along the walls of the void and the local chemistry of the cementitious matrix, sorbed Fe(III) ions at the void-cement interface undergo precipitation into thermodynamically favoured solid Fe(III) (hydr)oxides (Figure 4b). The simultaneous precipitation of Fe(III) (hydr)oxides along the walls of the void may further be catalysed by heterogeneous nucleation [40–45]. This hypothesis is supported by the fact that a thin homogeneous shell of corrosion products forms all along the void-cement interface in the early stages (Figure 3d). Its thickness or characteristics do not seem to vary significantly along this interface, as observed through SEM images (Figure 1) and XCT (Figure 2, Figure 3).

Specifically in the case of specimens exposed to wet and dry cycles (as in Figure 3c, d), one can hypothesise the precipitation of a uniform shell of corrosion products is also due to the influence of drying on the local activity of Fe(III) along the walls of the interfacial void. During drying or water desorption, Fe(III) ions witness a significant increase in their activity as the volume of water within the void reduces. As soon as the local concentration of Fe(III) along the walls of the void reaches saturation w.r.t. a solid Fe(III) (hydr)oxide, a uniform shell of solid corrosion products can be observed to precipitate.

*Inward growth of corrosion products*

It is worth highlighting that the type of Fe(III) (hydr)oxide that precipitates along the walls of the void depends on the degree of supersaturation achieved locally, the pH and the ageing time. For example, if the activity of Fe(III) is above the solubility of 2 line-ferrihydrite (thermodynamically least stable), the first precipitating solid Fe(III) (hydr)oxide phase (corrosion product) would be 2 line-ferrihydrite [46]. Depending on the pH of the electrolyte in the vicinity of the first layer of corrosion products, 2 line-ferrihydrite would transform to thermodynamically more stable phases such as goethite or haematite [46,47]. The transformation mechanism [46,48] would cause changes in the electrolyte chemistry (such as in the concentrations of Fe(III), OH$^-$, and possibly Fe(II)) locally, influencing the precipitation of subsequent layers of corrosion products (Figure 4b, c).

Over time, subsequent layers of iron oxides/hydroxides were observed to start precipitating along the boundary of the first layer of corrosion product formed at the void-cement interface (Figure 3f, h and represented through Figure 4c). This observation spans different samples exposed to a variety of conditions (as shown in Figure 1), and each layer of corrosion products seems to differ in chemistry when compared to the first layer (Figure 1). This can possibly be explained by the fact that aqueous Fe(II), Fe(III) and possibly mobile Fe(III) oxides/hydroxides present within the immobilised electrolyte adsorb onto the surface of the outer shell of corrosion product formed at the void-cement interface [42,49–52].

Previous work in the literature suggests that Fe(II) ions sorbed on the surfaces of ferric oxides can oxidise and form amorphous ferric oxides/hydroxides, which in turn could transform into magnetite [42,49–52]. It is also well known that Fe(III) ions can sorb strongly onto iron oxides and the sorbed Fe(III) may precipitate as amorphous iron oxides/hydroxides with a structure similar to that of the substrate (in this case, the first layer of corrosion products) [52]. This amorphous iron oxide/hydroxide may react with Fe(II) ions in the electrolyte to form magnetite [52]. While several studies have conducted ex-situ characterisation [30] of corrosion products within macroscopic interfacial voids and around the SCI, there is limited documentation concerning the characterisation of corrosion products in-situ. Relying on ex-situ characterisation for deciphering the mechanism governing precipitation of different corrosion products has its drawbacks [1,19], particularly when corrosion products are sensitive to exposure environments (eq. changes in $O_2$ concentration) and sample preparation (eq. cutting, polishing). Therefore, here we do not attempt to conclusively estimate the chemistry of different corrosion products precipitating within the void, and this needs to be considered for future research.

The formation of various corrosion products, including both the initial and subsequent layers, within macroscopic interfacial voids is significantly influenced by the chemical makeup of the solution within the void. The alkaline nature, with a pH similar to that of the pore solution, of the electrolyte within the interfacial void facilitates rapid oxidation of Fe(II) (predominantly $Fe(OH)_3^-$) into Fe(III) (predominantly $Fe(OH)_4^-$), and consequent precipitation of the Fe(III) (hydr)oxides [46]. The aqueous concentration and speciation of both Fe(II) and Fe(III) are dependent on the pH of the electrolyte and thermodynamic stability of different iron (hydr)oxides, and are kinetically controlled by the rate of anodic dissolution of the metal, the oxidation rate of Fe(II), the $E_{rev}^{Fe(II) \rightarrow Fe(III)}$ in solution, precipitation and transformation rate of Fe(III) (hydr)oxides. The solution within the void not only contains dissolved ferrous (below the solubility of Fe(II)) and ferric species but most likely also contains mobile Fe(III) oxides/hydroxides (possibly at the nano-meter scale [42]) that are possibly very different from those formed in the periphery of the void (Figure 4b). Therefore, particular emphasis must be laid on the prevailing chemistry of the solution within the macroscopic interfacial void and its influence on the precipitation of corrosion products and their inward growth.

To the authors' best understanding, this process of inward growth would continue until the entire interfacial void (under saturated conditions) is filled up with different corrosion products, or until a point where the corrosion products meets the entrapped air within the interfacial void (as in the case of drying), or when no electrolyte could reach the anodic site to sustain the transport of Fe ions due to the build-up of corrosion products within the corroding zone. But experimental evidence under controlled conditions needs to be acquired to confirm or deny the above hypotheses.

*Moisture within the void with corrosion products*

The lack of continuous experimental in-situ XCT measurements between the intial condition and the precipitation of the outer layer of corrosion products on the walls of interfacial voids (Figure 3b, d), does not allow us to decipher the moisture state within the void during the onset of corrosion. However, it is very likely that the onset of corrosion occurred when the macroscopic interfacial void was at least partially saturated with electrolyte [6]. The XCT measurements on samples exposed to wet and dry cycles conducted in this study show an interesting phenomenon, pertaining to the moisture state of macroscopic interfacial voids where corrosion products precipitate *vis-à-vis* other macroscopic voids at the SCI or in the bulk cementitious matrix.

A particularly interesting experimental observation was the following: once the first layer of corrosion products has precipitated along the walls of the void, in-situ XCT of the specimens (Figure 3c, d) show complete water saturation of the interfacial void (as observed with the lowest possible voxel size)[3]. As these measurements were conducted during the drying phase of the wet and dry cyclic exposure, it was observed that voids with corrosion products do not dry out as quickly as the other voids at the SCI or in the bulk cementitious matrix (Figure 3c, d). Observing full saturation of the void (while the other voids are filled by air – as shown in Figure 3c, d) suggests that the precipitation of the outer layer of corrosion products entraps the electrolyte within the void by decreasing its desorption rate during drying. The thickness of the outer layer of corrosion products is in the order of tens of μm, which could potentially generate such a sealing effect. The thickness of the outer layer of corrosion products is significantly smaller than the dimensions of the macroscopic voids. Therefore, it is highly improbable that the precipitation of the outer layer of corrosion products would reduce the volume of the void significantly. Such a negligible change in the void volume is thus not expected to affect the degree of water saturation to an appreciable extent. At later stages, the precipitation of corrosion products in the periphery of the interfacial void (Figure 3d) may induce microcracks that likely alter the microstructure of the corrosion products and cementitious matrix around the interfacial void and consequentially influence the water transport behaviour. Alternatively, corrosion products at the walls of the interfacial voids may lead to the blocking of some capillaries surrounding the void through which desorption takes place. However, the spatial resolution of the performed XCT acquisitions currently does not allow us to detect such phenomena and must be tackled in future studies. A similar situation of complete water saturation is observed when subsequent layers of corrosion products precipitate within the macroscopic interfacial void (Figure 3f).

---

[3] The presence of solution in the void is confirmed by the observations in Figure 3h, where prolonged drying of the specimen results in the complete replacement of water by air within the void.

### 3.1. Distribution of water in macroscopic voids at the SCI

For the hypothesised mechanism (Figure 4) involving processes such as oxidation, diffusion, hydrolysis, speciation, adsorption and precipitation/heterogeneous nucleation, occurring prior to the precipitation of corrosion products to be true, the presence of an aqueous phase within the void is essential. Fagerlund [53,54] proposed the filling/adsorption of water in the over-capillary range pores to be a result of the gradual dissolution of air in the pore water due to the over-pressure created by the saturation of capillary pores. As the solubility of air is proportional to the over-pressure, Fagerlund [53,54] showed that smaller pores fill more readily with more water than coarser pores. Based on this mechanism and several assumptions on the involved parameters (e.g. gas diffusivity), the minimum time for coarse air voids above 0.5 mm to achieve a fully saturated state was estimated to be several years or even decades. Several studies [55–58] have shown that the degree of capillary saturation is almost never 100 %, leading to the assumption that coarse air voids must only contain adsorbed water along their walls [1]. Data from several structures have shown contrarian evidence [55,56,59], where some coarse air voids within the structure can be saturated to a great extent even in the absence of hydrostatic over-pressure. Therefore, it can be presumed that macroscopic voids at the SCI can host a variety of moisture conditions, some partially filled with water and some with a thin adsorbed film of water or completely air filled. However, to the authors' best knowledge there does not exist any evidence pertaining to the distribution of water within the macroscopic interfacial void and whether the distribution of water within the void influences the manner in which corrosion products precipitate within the void.

It is challenging to experimentally visualise the ingress of water within macroscopic interfacial voids. We used XCT[4] to shed light on the distribution of water within macroscopic voids at the SCI in reinforced mortar. We assessed the distribution of water within macroscopic voids at the SCI in reinforced mortar specimens submerged for 8 weeks. The acquired 2D XCT images

---

[4] Experimental details and methodology of XCT of reinforced concrete specimens can be found in the Supplementary Note 1.

(Figure 5) show that the air-liquid configuration in macroscopic voids at the SCI can significantly vary. In the tested specimen, at least four interfacial voids are present at the SCI after casting (Figure 5a), out of which three are of diameters in the mm range. After 8 weeks of submersion in NaCl solution, the coarse interfacial voids (with a diameter of ~1 mm) are partially saturated (Figure 5b), similar to observations by Jacobson [59] and Relling [56,57]. Among the partially saturated macroscopic voids at the SCI (Figure 5b), the gaseous phase is detached from the steel in two out of three voids, whereas in one case the steel is in direct contact with the gaseous phase.

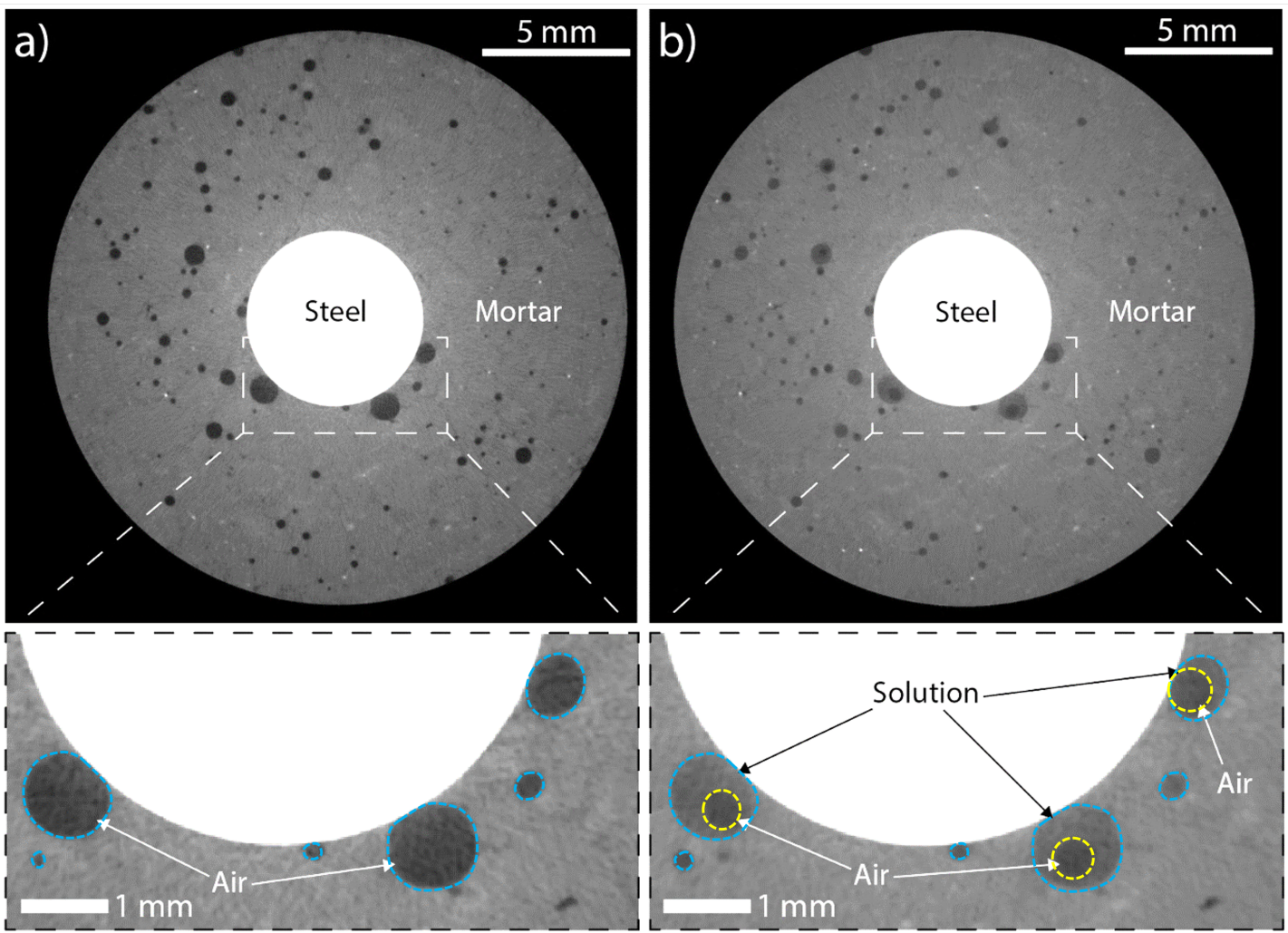

Figure 5: X-ray CT 2D cross sections of voids at the SCI: a) reinforced mortar specimen in its undisturbed state (after hydration); b) reinforced mortar specimen after 8 weeks of submersion in 3.5 wt.% NaCl solution, indicating the gaseous and liquid phases in coarse interfacial voids with white and black arrows, respectively. Blue dashed lines indicate the boundaries of the macroscopic void and the area within the yellow dashed lines in (b) represents the air bubble within a partially saturated macroscopic interfacial void.

As proposed by Fagerlund [53,54], the sub-micron scale microstructure surrounding the macroscopic void influences how water fills and distributes within the void. The XCT data shown in Figure 5 is limited by both, the spatial and temporal resolution required to decipher why moisture distributes differently in macroscopic voids of similar sizes at the SCI. Therefore, we here use computational fluid dynamics (CFD)[5] to simulate the fully resolved multiphase flow of air-water phases at the pore scale to understand the influence of certain parameters (e.g., the presence of a cementitious layer between the steel and the interfacial void) on the distribution of water within the void.

To do this, we consider the macroscopic interfacial void to be surrounded by a number of capillaries through which water can ingress into the void, all represented by a single large capillary denoted as Pore A (Figure 6). We consider the macroscopic interfacial void to be also surrounded by a number of outlet capillaries through which fluid, either water or air, can flow out from the void. These are represented by Pore B and Pore C. In several regions at the SCI, a thin layer of cement paste may be present between the macroscopic void and the steel, which would alter the wettability or contact angle at the interface of the void/steel. We study this by altering the contact angle at the void/steel interface from 75° (common for water-steel-air) to 0° (as would be expected for water-cement-air). These two cases are represented as Case 1 (contact angle: 75°) and Case 2 (contact angle: 0°) in Figure 6a-e and Figure 6f-j, respectively. The results shown in Figure 6 illustrate the evolution of the air-liquid distribution once water has entered the macroscopic void. Using these simulations, we can provide a possible explanation for the different air-liquid distributions, either the air bubble being attached to or detached from the steel surface, as observed in the XCT experiments (shown in Figure 5b). To understand how water ingresses into the void and attains the configurations shown in Figure 6a and Figure 6f, the reader is directed to Supplementary Note 2.

---

[5] Methodology details concerning computational fluid dynamics are discussed in detail in Supplementary Note 2

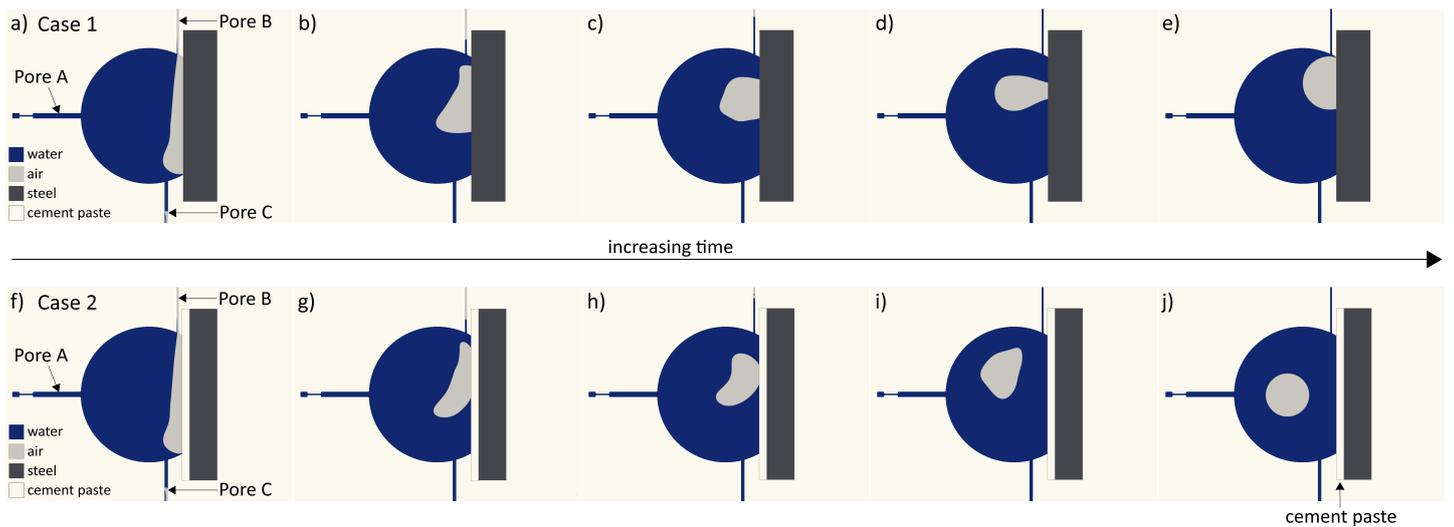

Figure 6: CFD simulation results of the interfacial void-filling process without (a-e) and with (f-j) the presence of a thin cement paste layer covering the steel surface. Figures 6a-d and Figures 6f-i show the time evolution (in the order of tens of milliseconds) of air entrapment before attaining the final states (e) and (j). Simulations indicate that the presence of the cement paste layer covering the steel surface could be one of the reasons for the detachment of the air bubble from the steel. See Supplementary Note 2 for methodology.

Once water within the void has taken up the configuration as shown in Figure S 1e (Supplementary Note 2), water advances rapidly in the outlet pore (Pore C) as a consequence of the narrow geometry of the pore (Pore C) and large capillary pressure (Figure 6a, f). The evolution of the shape of the meniscus and the distribution of air and water inside the macroscopic void, are both strongly influenced by the absence (Figure 6a-e) or presence of a thin cement paste layer (Figure 6f-j) on the surface of the steel. The meniscus inside the void undergoes rapid dynamic reconfigurations due to the strong interfacial tension between air and liquid (Figure 6b-c and Figure 6g-h). Moreover, the dynamic reconfiguration of the air-liquid interface inside the void caused by the filling of Pore C is strong enough to overcome the geometrical barrier at the mouth of Pore B (Figure 6b and Figure 6g), which leads to the filling of Pore B as well. Once both outlet capillaries get filled by liquid, the air phase becomes discontinuous and is trapped within the void volume, oscillating (Figure 6b-d and Figure 6g-i) until it achieves a constant curvature, representing equilibrium conditions (Figure 6e and Figure 6j).

Since the transport of water is strongly influenced by the wettability of the surrounding surfaces (defined by their contact angle values), having a thin cement paste layer between the steel and the void may thus drastically modify the distribution of water in interfacial voids (compare Figure 6d, e and Figure 6i, j). After water starts filling the outlet capillaries, its attraction to the cement paste layer is higher (compared to Case 1 where water was in direct contact with steel Figure 6d, e) and, together with interfacial tension effects, leads to detachment of the air phase from the steel surface. Again, the air bubble undergoes dynamic reconfiguration and oscillates until it reaches its equilibrium (spherical) shape (Figure 6j). Once the trapped air bubble detaches from the solid surface, it is not likely that it will re-establish contact with the solid steel surface (like in Case 1 - Figure 6e) because of the high wettability of the cement paste layer. In both scenarios, Case 1 and Case 2, interfacial tension on the gas-liquid interface results from the imbalance between cohesive and adhesive forces at the molecular level. Such forces tend to minimize the energy of the system by minimizing gas-liquid interfacial surface area upon striving towards equilibrium, which results in either air bubbles (somewhat spherical in shape, actual spherical radius depending on the contact angle) attached to the steel surface (Case 1 - Figure 6e) or spherical air bubbles enclosed within the water in the void (Case 2 - Figure 6j).

It is worthwhile to consider that prior to any corrosion initiation, completely saturated macroscopic voids (with a diameter in the range of a few mm) are rarely encountered under real exposure conditions and concretes, whereas partially saturated voids at the SCI with different air-water configurations can be considered as commonplace in practice (Section 3.1). Interestingly, we clearly observe that the distribution of water within the macroscopic void at the SCI, as experimentally observed (Figure 5) and plausibly explained (Figure 6), exhibits a shape similar to that of corrosion products precipitating at the walls of the void. This strongly suggests that the distribution of water within the macroscopic interfacial void is closely linked to the processes pertaining to the precipitation of corrosion products, as highlighted above in Section 3.

The results from the above simulations provide insights into some of the factors that may influence the distribution of water and air within the macroscopic void at the SCI differ, as observed through XCT in Figure 5b. These results suggest that the characteristics of the capillary porous network surrounding the interfacial void and the presence or absence of a thin cement paste layer on the steel surface influence the distribution of air and water in these voids. It must be recognised that though these simulations (Figure 6) represent the conditions prevailing within and around the macroscopic interfacial void simplistically, they exemplify

CFD as a useful and powerful tool to provide mechanistic insights into processes that are currently not possible to experimentally observe due to limitations in spatial and temporal resolution. Future research should be directed towards modelling the ingress of water (at the pore-scale) into interfacial voids at the SCI by considering several complexities such as the interconnectivity and geometry of the surrounding capillary porous network (a preliminary example of this can be found in Supplementary Note 2), the local changes in the wettability (or contact angles) due to the precipitation of different corrosion products, and the microstructure of the corrosion products.

## 4. Implication on stress development within macroscopic interfacial voids

The precipitation of corrosion products within macroscopic interfacial voids leads to the development of stresses in and around the voids. These induced stresses may cause micro-cracking, which can potentially influence all the processes discussed above. The process by which a precipitating solid induces stress in a confined space is comprehensively understood in the context of capillary pores, which vary in size from nanometers to micrometers. According to crystallization theory, a precipitate must crystallize, assimilate, and achieve a certain concentration threshold to apply pressure onto the capillary pore walls [60–63]. At these length scales, the expansion of the precipitate progresses from the interior of the void to its walls. Conversely, in larger interfacial voids, deposition of corrosion products first initiates at the walls (due to adsorption and heterogeneous crystallisation) of the void, as illustrated in Figure 1, Figure 2 and Figure 3. Additionally, corrosion products deposit in successive layers around the walls of the void where layers may be chemically different from each other (Figure 1). These observations raise the question of how such precipitation patterns apply pressure to the walls of the void, initiating and propagating micro-cracks (refer to cracks in Figure 2a).

It is essential to recognise that corrosion products precipitate both within macroscopic interfacial voids and concurrently within the capillary pores of the surrounding concrete matrix. Our focus is not to compare the importance or magnitude of stresses generated by the precipitation of corrosion products in capillary pores and macroscopic interfacial voids in the context of corrosion-driven fracture. Instead, we intend to lay emphasis on how stresses develop when corrosion products precipitate along the walls of macroscopic voids and how this process may influence the microstructure of the surrounding concrete matrix.

It is often assumed in the literature [64–66] that in large interfacial voids, the development of pressure and subsequent crack initiation results from the isotropic volumetric expansion of the precipitate (as shown in Supplementary Note 3). Nonetheless, the above findings suggest that application of pressure in macroscopic voids is much more complex and could involve an intricate interplay between various elements. Notably, the existence of chemically distinct layers of corrosion products, varying in relative density (w.r.t. the pore solution containing Fe(III) (aq.)) and elastic modulus, indicates that interactions among these layers are necessary to understand how stress develops within and around the macroscopic interfacial voids. For instance, if a layer of corrosion products is surrounded by a different corrosion product with different elastic modulus and expansion coefficients, then in such a scenario, how stress develops within the surrounding medium depends on how (both in magnitude and in time) the pressure due to the expansion of the inner layer is passed through the surrounding layer to the pore walls.

Furthermore, as previously mentioned, in contrast to the capillary process (which is consistently considered to be saturated with the concrete pore solution), the degree of saturation in the macroscopic interfacial voids varies depending on environmental conditions [67] and the surrounding capillary porous network. Figure 3c depicts a layer of corrosion products forming along the walls of unsaturated interfacial voids. It remains uncertain if such corrosion products can apply pressure and thereby, play a role in causing fractures.

In order to get a first-hand insight into this interplay, we employ numerical simulations[6] to simulate the stress states within the surrounding cementitious matrix as well as within the corrosion product layers due to their precipitation. Figure 7 shows the stress states for 2 scenarios: (a) when a single layer of corrosion product layer forms from solution on the walls of a saturated macroscopic voids, and (b) when two distinct layers of corrosion products form from solution within a saturated macroscopic void (like the XCT observations in Figure 3). For reference, we also include the case where corrosion products form and grow in a conventional manner, *i.e.*, from inside to outside (Supplementary Note 3). In the scenario of a single corrosion product layer precipitating on the walls of the macroscopic void, our study suggests that as this layer precipitates from solution, the major principal stresses (maximum normal stress experienced by a point within a material along the plane where shear stress is 0 [68])

---

[6] Methodology of the numerical simulations for computing the stress states are discussed in Supplementary Note 3

within the surrounding cementitious matrix can still become considerably large enough to cause cracking (refer to Figure 7a). This can radily cause fracture initiation within the cementitious matrix, similar to the case when the growth of corrosion products occurs isometrically (see Figure S3b).

In the case of the formation of two layers of corrosion product within the macroscopic interfacial voids, we see that the stress state within the cement phase significantly depends on the material properties of the precipitated corrosion products (Figure 7b). As the two layers precipitate, a middle layer (see Figure 7b – black line) with a lower Young's modulus can delay the time to achieve the stress state required for crack initiation of the surrounding cementitious matrix. Conversely, a middle layer with a stiffer corrosion product (see Figure 7b – red line) could accelerate the attainment of the critical stress state (crack initiation) within the surrounding cementitious matrix. As mentioned earlier, experimental observations have suggested that these layers are chemically distinct (goethite, magnetite, ferrihydrite) [29] with significantly different Young's modules [30]. Furthermore, these corrosion products have different relative densities (much greater than the pore solution containing Fe(III) (aq.)), which could further complex the development of stresses. One significant difference from the previous scenario (Figure 7a) is that the major principal stresses within the corrosion product layers can also become sufficiently large (see Figure 7b). This can readily result in fracture initiation within the corrosion product layer, which could be why we experimentally observe micro-cracks within these precipitates (as shown in Figure 1, Figure 2, Figure 3). Such cracks are significant to the corrosion propagation process as they can potentially create new pathways for the pore solution or moisture to infiltrate the interfacial void and initiate further precipitation (formation of a second corrosion product layer).

Our observations are in stark contrast with the conventional approach of precipitate growth within fully-filled pores and subsequent stress development in the surrounding medium (see Supplementary Note 3). These observations suggest that a macroscopic void does not have to be completely filled (conventional approach) with corrosion products in order to exert pressure onto the walls of macroscopic voids. We consider these findings noteworthy as they demonstrate that the precipitation of corrosion products within the macroscopic voids can alter the microstructure around them. This alteration can significantly impact various corrosion-induced mechanisms, highlighting the need for further investigation.

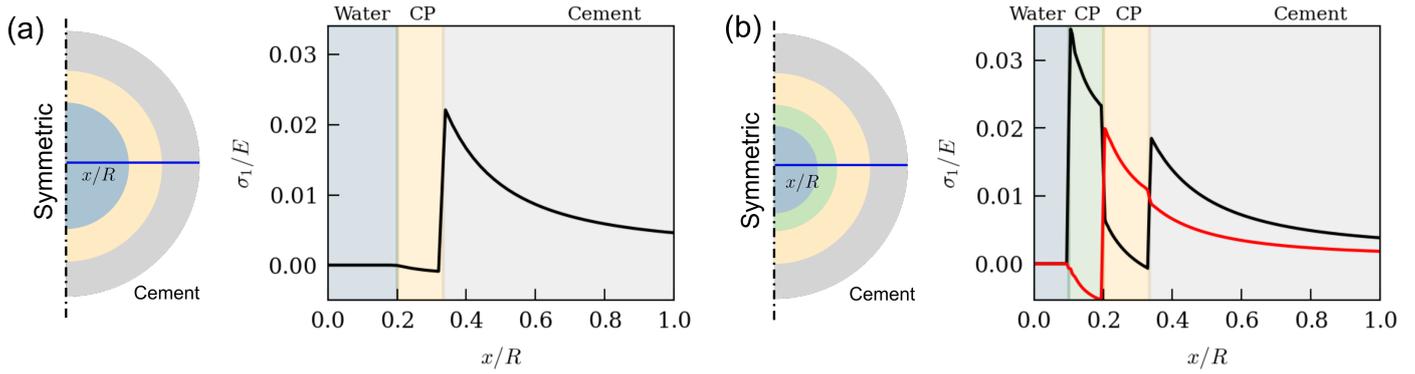

Figure 7: Numerical simulation results (line plots) showing the major principal stresses within different corrosion products and the surrounding cementitious matrix along the cross section (shown in blue). (a) When a single layer of corrosion product layer precipitates (denoted as CP in the figure) from solution on the walls of a saturated macroscopic interfacial void, and (b) when two layers of different corrosion products (CP) precipitate on the walls of a saturated macroscopic interfacial void. In Figure 7b, the red line refers to the case when the middle layer of corrosion product is softer than the outer layer, whereas the black line refers to the case when the outer layer of corrosion product is softer than the middle layer, The y-axis values show the major principal stresses normalised w.r.t. the Young's modulus of the respective material, and the x-axis is the normalized distance from the centre of the void.

## Conclusion and future research needs

Through a combination of in-situ X-ray computed tomography (XCT) experiments, numerical simulations of water uptake, and theoretical thermodynamic and kinetic analyses, we have proposed plausible hypotheses to explain the processes governing the precipitation of corrosion products within macroscopic voids at the steel-concrete interface (SCI). Our findings suggest that the precipitation of corrosion products as 'concentric shells' along the void walls is primarily driven by the preferential sorption of aqueous Fe(III) onto these walls, followed by the heterogeneous crystallization of Fe(III). This is corroborated by the observed resemblance between the water distribution within the voids and the morphology of the precipitated corrosion products.

The chemical composition of the Fe(III) (hydr)oxide layers is influenced by a range of interdependent factors, including the local degree of supersaturation, pH, Fe(II) oxidation rate, and the ageing time of the layers. A particularly intriguing observation from our experiments is the differences in desorption behavior in voids with and without a shell of corrosion products. We propose that the formation of corrosion products along the void walls may induce stresses both within the precipitated layers and in the surrounding cementitious matrix. The magnitude of these stresses, which depends on the material properties of the corrosion products and the surrounding cementitious matrix, could eventually (at later stages) become sufficient to initiate cracking. Such cracking has the potential to create new pathways, thereby influencing subsequent water uptake, desorption processes, and the propagation of corrosion.

However, we believe that substantial progress is needed in terms of the scientific and inter-disciplinary understanding of all the coupled processes relevant for precipitation of corrosion products and corrosion-related damage. Key areas that require particular attention include:

– *Corrosion initiation:* There is an urgent need to develop a fundamental understanding on whether there exists a relationship between the degree of water saturation of macroscopic interfacial voids and the probability of corrosion initiation. Additionally, characterising the moisture state in macroscopic interfacial voids shortly before and after corrosion initiation would provide valuable insight into its influence on corrosion initiation. In this regards, it is also vital to develop a thorough understanding of the influence of different parameters - the presence of a cement paste layer between the void and the steel, exposure conditions, the microstructure of the surrounding concrete matrix - on the distribution of water and air within the macroscopic void, and consequentially on corrosion initiation.

– *Water ingress into macroscopic voids:* The presence of water and its distribution within macroscopic voids is critical for both corrosion and precipitation of corrosion products. Here, we exemplify some of the influential parameters, such as wettability of different surfaces at the SCI and the interconnectivity of the capillary pores, to explain the differences in the distribution of water and air within similar sized macroscopic voids at the SCI. However, additional work is needed to comprehensively and fundamentally understand water ingress within macroscopic voids. As highlighted in Section 3.1, important factors such as the microstructure of the concrete matrix surrounding the macroscopic voids – interconnectivity of the capillary pores, the exposure conditions, the changes in the local wettability (or contact angle) due to precipitation of different corrosion products, condensation, drying and the microstructure of the corrosion products, among other, should be considered in future studies.

- *Precipitation of corrosion products:* While there exist several studies pertaining to the homogeneous and heterogeneous oxidation rate of Fe(II) in near-neutral conditions, kinetic data for both homogeneous and heterogeneous oxidation of Fe(II) in alkaline electrolytes is non-existent. To understand the precipitation of different Fe(III) corrosion products in alkaline cementitious materials, knowing the oxidation rate of Fe(II) is paramount. Additionally, significant work is required in the field of studying the adsorption behaviour of Fe(II) and Fe(III) on cementitious phases and different iron (hydr)oxides to understand the growth of corrosion products. Here, particular attention needs to be directed towards characterising the local electrolyte chemistry prevailing at the interface of different iron (hydr)oxides as well as in the corroding zone.

Though corrosion products (chemically distinct) precipitate around the walls of the void, we currently do not know which corrosion products precipitate and how do they evolve over time as a function of the electrolyte chemistry within the void. Therefore, characterising the different corrosion products may be particularly useful in this context. This would help us in filling some of the gaps in the literature as to which corrosion products precipitate in cementitious environments and how do they transform. This information is critical to develop models to fundamentally understand the stresses generated within corrosion products in macroscopic voids and the surrounding matrix in concrete structures or in other porous media. Additionally, experimental efforts must be made to validate the development of cracks in the surrounding concrete matrix and the corrosion products, as predicted by the numerical simulation presented here.

Addressing these gaps will certainly enhance our understanding of corrosion mechanisms and improve durability predictions for reinforced concrete structures.

## Acknowledgements

The authors would like to thank the Swiss National Science Foundation (PP00P2_194812 and CRSII5_205883) and European Research Council (ERC Strating Grant: Taming Corrosion 848794). The XCT measurements were done at the Paul Scherrer Intsitut ICON beamline.

Thanks are due to Susanna Governo for preparing the samples, discussions and helping with conducting XCT measurements, and to the beamline staff, particularly, Dr Mahdieh Shakoorioskooie.

**Funding**

Swiss National Science Foundation (PP00P2_194812 and CRSII5_205883) and European Research Council (ERC Strating Grant: Taming Corrosion 848794).

**Conflict of Interest**

The authors declare no conflict of interest.